\documentclass[twocolumn]{aastex631}
\usepackage{amsmath}
\usepackage{graphicx}

\graphicspath{ {figures/} }

\newcommand{\unwisetwomass}[0]{unWISE-2MASS }
\newcommand{\vx}{\vec{x}}

\newcommand{\fcorr}{\(f_{\rm corr}\)}

\begin{document}
\title{Constraints on the Correlation of IceCube Neutrinos with Tracers of Large-Scale Structure}

\affiliation{III. Physikalisches Institut, RWTH Aachen University, D-52056 Aachen, Germany}
\affiliation{Department of Physics, University of Adelaide, Adelaide, 5005, Australia}
\affiliation{Dept. of Physics and Astronomy, University of Alaska Anchorage, 3211 Providence Dr., Anchorage, AK 99508, USA}
\affiliation{School of Physics and Center for Relativistic Astrophysics, Georgia Institute of Technology, Atlanta, GA 30332, USA}
\affiliation{Dept. of Physics, Southern University, Baton Rouge, LA 70813, USA}
\affiliation{Dept. of Physics, University of California, Berkeley, CA 94720, USA}
\affiliation{Lawrence Berkeley National Laboratory, Berkeley, CA 94720, USA}
\affiliation{Institut f{\"u}r Physik, Humboldt-Universit{\"a}t zu Berlin, D-12489 Berlin, Germany}
\affiliation{Fakult{\"a}t f{\"u}r Physik {\&} Astronomie, Ruhr-Universit{\"a}t Bochum, D-44780 Bochum, Germany}
\affiliation{Universit{\'e} Libre de Bruxelles, Science Faculty CP230, B-1050 Brussels, Belgium}
\affiliation{Vrije Universiteit Brussel (VUB), Dienst ELEM, B-1050 Brussels, Belgium}
\affiliation{Dept. of Physics, Simon Fraser University, Burnaby, BC V5A 1S6, Canada}
\affiliation{Department of Physics and Laboratory for Particle Physics and Cosmology, Harvard University, Cambridge, MA 02138, USA}
\affiliation{Dept. of Physics, Massachusetts Institute of Technology, Cambridge, MA 02139, USA}
\affiliation{Dept. of Physics and The International Center for Hadron Astrophysics, Chiba University, Chiba 263-8522, Japan}
\affiliation{Department of Physics, Loyola University Chicago, Chicago, IL 60660, USA}
\affiliation{Dept. of Physics and Astronomy, University of Canterbury, Private Bag 4800, Christchurch, New Zealand}
\affiliation{Dept. of Physics, University of Maryland, College Park, MD 20742, USA}
\affiliation{Dept. of Astronomy, Ohio State University, Columbus, OH 43210, USA}
\affiliation{Dept. of Physics and Center for Cosmology and Astro-Particle Physics, Ohio State University, Columbus, OH 43210, USA}
\affiliation{Niels Bohr Institute, University of Copenhagen, DK-2100 Copenhagen, Denmark}
\affiliation{Dept. of Physics, TU Dortmund University, D-44221 Dortmund, Germany}
\affiliation{Dept. of Physics and Astronomy, Michigan State University, East Lansing, MI 48824, USA}
\affiliation{Dept. of Physics, University of Alberta, Edmonton, Alberta, T6G 2E1, Canada}
\affiliation{Erlangen Centre for Astroparticle Physics, Friedrich-Alexander-Universit{\"a}t Erlangen-N{\"u}rnberg, D-91058 Erlangen, Germany}
\affiliation{Physik-department, Technische Universit{\"a}t M{\"u}nchen, D-85748 Garching, Germany}
\affiliation{D{\'e}partement de physique nucl{\'e}aire et corpusculaire, Universit{\'e} de Gen{\`e}ve, CH-1211 Gen{\`e}ve, Switzerland}
\affiliation{Dept. of Physics and Astronomy, University of Gent, B-9000 Gent, Belgium}
\affiliation{Dept. of Physics and Astronomy, University of California, Irvine, CA 92697, USA}
\affiliation{Karlsruhe Institute of Technology, Institute for Astroparticle Physics, D-76021 Karlsruhe, Germany}
\affiliation{Karlsruhe Institute of Technology, Institute of Experimental Particle Physics, D-76021 Karlsruhe, Germany}
\affiliation{Dept. of Physics, Engineering Physics, and Astronomy, Queen's University, Kingston, ON K7L 3N6, Canada}
\affiliation{Department of Physics {\&} Astronomy, University of Nevada, Las Vegas, NV 89154, USA}
\affiliation{Nevada Center for Astrophysics, University of Nevada, Las Vegas, NV 89154, USA}
\affiliation{Dept. of Physics and Astronomy, University of Kansas, Lawrence, KS 66045, USA}
\affiliation{Centre for Cosmology, Particle Physics and Phenomenology - CP3, Universit{\'e} catholique de Louvain, Louvain-la-Neuve, Belgium}
\affiliation{Department of Physics, Mercer University, Macon, GA 31207-0001, USA}
\affiliation{Dept. of Astronomy, University of Wisconsin{\textemdash}Madison, Madison, WI 53706, USA}
\affiliation{Dept. of Physics and Wisconsin IceCube Particle Astrophysics Center, University of Wisconsin{\textemdash}Madison, Madison, WI 53706, USA}
\affiliation{Institute of Physics, University of Mainz, Staudinger Weg 7, D-55099 Mainz, Germany}
\affiliation{Department of Physics, Marquette University, Milwaukee, WI 53201, USA}
\affiliation{Institut f{\"u}r Kernphysik, Universit{\"a}t M{\"u}nster, D-48149 M{\"u}nster, Germany}
\affiliation{Bartol Research Institute and Dept. of Physics and Astronomy, University of Delaware, Newark, DE 19716, USA}
\affiliation{Dept. of Physics, Yale University, New Haven, CT 06520, USA}
\affiliation{Columbia Astrophysics and Nevis Laboratories, Columbia University, New York, NY 10027, USA}
\affiliation{Dept. of Physics, University of Oxford, Parks Road, Oxford OX1 3PU, United Kingdom}
\affiliation{Dipartimento di Fisica e Astronomia Galileo Galilei, Universit{\`a} Degli Studi di Padova, I-35122 Padova PD, Italy}
\affiliation{Dept. of Physics, Drexel University, 3141 Chestnut Street, Philadelphia, PA 19104, USA}
\affiliation{Physics Department, South Dakota School of Mines and Technology, Rapid City, SD 57701, USA}
\affiliation{Dept. of Physics, University of Wisconsin, River Falls, WI 54022, USA}
\affiliation{Dept. of Physics and Astronomy, University of Rochester, Rochester, NY 14627, USA}
\affiliation{Department of Physics and Astronomy, University of Utah, Salt Lake City, UT 84112, USA}
\affiliation{Dept. of Physics, Chung-Ang University, Seoul 06974, Republic of Korea}
\affiliation{Oskar Klein Centre and Dept. of Physics, Stockholm University, SE-10691 Stockholm, Sweden}
\affiliation{Dept. of Physics and Astronomy, Stony Brook University, Stony Brook, NY 11794-3800, USA}
\affiliation{Dept. of Physics, Sungkyunkwan University, Suwon 16419, Republic of Korea}
\affiliation{Institute of Physics, Academia Sinica, Taipei, 11529, Taiwan}
\affiliation{Dept. of Physics and Astronomy, University of Alabama, Tuscaloosa, AL 35487, USA}
\affiliation{Dept. of Astronomy and Astrophysics, Pennsylvania State University, University Park, PA 16802, USA}
\affiliation{Dept. of Physics, Pennsylvania State University, University Park, PA 16802, USA}
\affiliation{Dept. of Physics and Astronomy, Uppsala University, Box 516, SE-75120 Uppsala, Sweden}
\affiliation{Dept. of Physics, University of Wuppertal, D-42119 Wuppertal, Germany}
\affiliation{Deutsches Elektronen-Synchrotron DESY, Platanenallee 6, D-15738 Zeuthen, Germany}

\author[0000-0001-6141-4205]{R. Abbasi}
\affiliation{Department of Physics, Loyola University Chicago, Chicago, IL 60660, USA}

\author[0000-0001-8952-588X]{M. Ackermann}
\affiliation{Deutsches Elektronen-Synchrotron DESY, Platanenallee 6, D-15738 Zeuthen, Germany}

\author{J. Adams}
\affiliation{Dept. of Physics and Astronomy, University of Canterbury, Private Bag 4800, Christchurch, New Zealand}

\author[0000-0002-9714-8866]{S. K. Agarwalla}
\altaffiliation{also at Institute of Physics, Sachivalaya Marg, Sainik School Post, Bhubaneswar 751005, India}
\affiliation{Dept. of Physics and Wisconsin IceCube Particle Astrophysics Center, University of Wisconsin{\textemdash}Madison, Madison, WI 53706, USA}

\author[0000-0003-2252-9514]{J. A. Aguilar}
\affiliation{Universit{\'e} Libre de Bruxelles, Science Faculty CP230, B-1050 Brussels, Belgium}

\author[0000-0003-0709-5631]{M. Ahlers}
\affiliation{Niels Bohr Institute, University of Copenhagen, DK-2100 Copenhagen, Denmark}

\author[0000-0002-9534-9189]{J.M. Alameddine}
\affiliation{Dept. of Physics, TU Dortmund University, D-44221 Dortmund, Germany}

\author[0009-0001-2444-4162]{S. Ali}
\affiliation{Dept. of Physics and Astronomy, University of Kansas, Lawrence, KS 66045, USA}

\author{N. M. Amin}
\affiliation{Bartol Research Institute and Dept. of Physics and Astronomy, University of Delaware, Newark, DE 19716, USA}

\author[0000-0001-9394-0007]{K. Andeen}
\affiliation{Department of Physics, Marquette University, Milwaukee, WI 53201, USA}

% ...

\date{\today}


\email{analysis@icecube.wisc.edu}

\collaboration{432}{IceCube Collaboration}

\begin{abstract}

The IceCube Neutrino Observatory has observed extragalactic astrophysical neutrinos with an apparently isotropic distribution. Only a small fraction of the observed astrophysical neutrinos can be explained by known sources. Neutrino production is thought to occur in energetic environments that are ultimately powered by the gravitational collapse of dense regions of the large-scale mass distribution in the universe. 
Whatever their identity, neutrino sources likely trace this large-scale mass distribution. 
The clustering of neutrinos with a tracer of the large-scale structure may provide insight into the distribution of neutrino sources with respect to redshift and the identity of neutrino sources. 
We implement a two-point angular cross-correlation of the Northern sky track events with an infrared galaxy catalog derived from WISE and 2MASS source catalogs that trace the nearby large-scale structure. 
No statistically significant correlation is found between the neutrinos and this infrared galaxy catalog. 
We find that $\leq ~$54\% of the diffuse muon neutrino flux can be attributed to sources correlated with the galaxy catalog with 90\% confidence.
Additionally, when assuming that the neutrino source comoving density evolves following a power-law in redshift, $dN_s/dV \propto (1+z)^{k}$, we find that sources with negative evolution, in particular $k < -1.75$, are disfavored at the 90\% confidence level. 

\end{abstract}


\section{Introduction} \label{sec:intro}

The IceCube Neutrino Observatory has observed a diffuse flux of astrophysical neutrinos with extragalactic origin \citep{icecubecollaborationEvidenceHighEnergyExtraterrestrial2013, icecubecollaborationEvidenceAstrophysicalMuon2015, icecubecollaborationMeasurementsUsingInelasticity2019, icecubecollaborationCharacteristicsDiffuseAstrophysical2020,icecubecollaborationIceCubeHighenergyStarting2021,  abbasiImprovedCharacterizationAstrophysical2022a, abbasiCharacterizationAstrophysicalDiffuse2024}.
For virtually all of these extragalactic astrophysical neutrinos, their sources remain unknown.
The extragalactic diffuse astrophysical neutrino flux appears isotropic; no large-scale anisotropy has been found.
Point-like neutrino emission has been observed from the Seyfert galaxy NGC 1068 and the blazar TXS 0506+056 \citep{icecubecollaborationEvidenceNeutrinoEmission2022, theicecubecollaborationMultimessengerObservationsFlaring2018, icecubecollaborationNeutrinoEmissionDirection2018}; however, these two sources can only explain a few percent of the diffuse neutrino flux.
Astrophysical neutrinos have been observed from the galactic plane \citep{icecubecollaboration*+ObservationHighenergyNeutrinos2023}, but they cannot account for the entire astrophysical diffuse flux.

There have been several prior studies searching for anisotropies in the IceCube neutrino data.
\citet{fangCrosscorrelationStudyHighenergy2020} searched for correlations of 3 years of IceCube data with infrared galaxies and placed an upper limit on the correlation.
\citet{aartsenConstraintsNeutrinoEmission2020} searched for neutrino emission correlated with the large-scale structure with both \(z<0.1\) and \(z<0.03\) using the 2MASS Redshift Survey and a template-based search with 7 years of muon neutrino events.
The upper limits on this neutrino emission are consistent with the observed diffuse flux.
\citet{negroCrosscorrelationStudyIceCube2023} performed a two-point angular cross-correlation analysis of IceCube neutrinos with the anisotropic unresolved $\gamma$-ray background (UGRB) observed by Fermi-LAT \citep{ackermannUnresolvedGammaRaySky2018}.
They found that less than 1\% of the observed diffuse flux could be explained by the decay of pions created in energetic unresolved blazars in the {\it Fermi}-LAT \(\gamma\)-ray background.
\citet{ouelletteCrosscorrelatingIceCubeNeutrinos2024b} performed a cross-correlation analysis across multiple redshift bins, resulting in upper limits on the correlation of IceCube neutrinos and the large-scale structure over a range of redshifts.

The two-point angular cross-correlation function is widely used in cosmology to measure the degree of clustering of two astrophysical populations \citep{peeblesLargescaleStructureUniverse1980a}.
The two-point angular cross-power spectrum is the correlation function expressed in the basis of spherical harmonics.
The cross-power spectrum can be calculated quickly for large catalogs when pixelized, so it is computationally efficient for the analysis of millions of neutrinos with a similarly sized tracer of large-scale structure.
We optimize the method of \citet{fangCrosscorrelationStudyHighenergy2020} and apply the two-point angular cross-correlation function to a set of Northern sky muon neutrino candidates observed by IceCube from 2011 - 2021 and an infrared galaxy catalog. 

We explain the derivation of the galaxy catalog in Section~\ref{sec:cat}. We describe the IceCube data used for this analysis in Section~\ref{sec:IceCube} and the analysis method in Section~\ref{sec:method}. We present the results in Section~\ref{sec:results}. We discuss the results in Section~\ref{sec:dis} and conclude in Section~\ref{sec:conclude}.

\section{unWISE-2MASS Galaxy Catalog}\label{sec:cat}

\begin{figure*}[t]
    \centering
    \plottwo{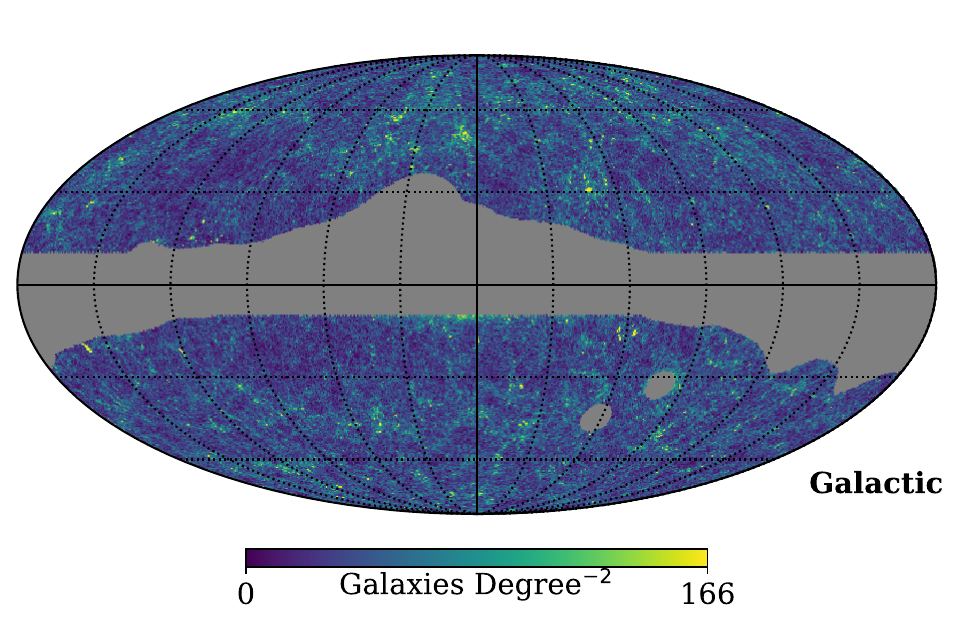}{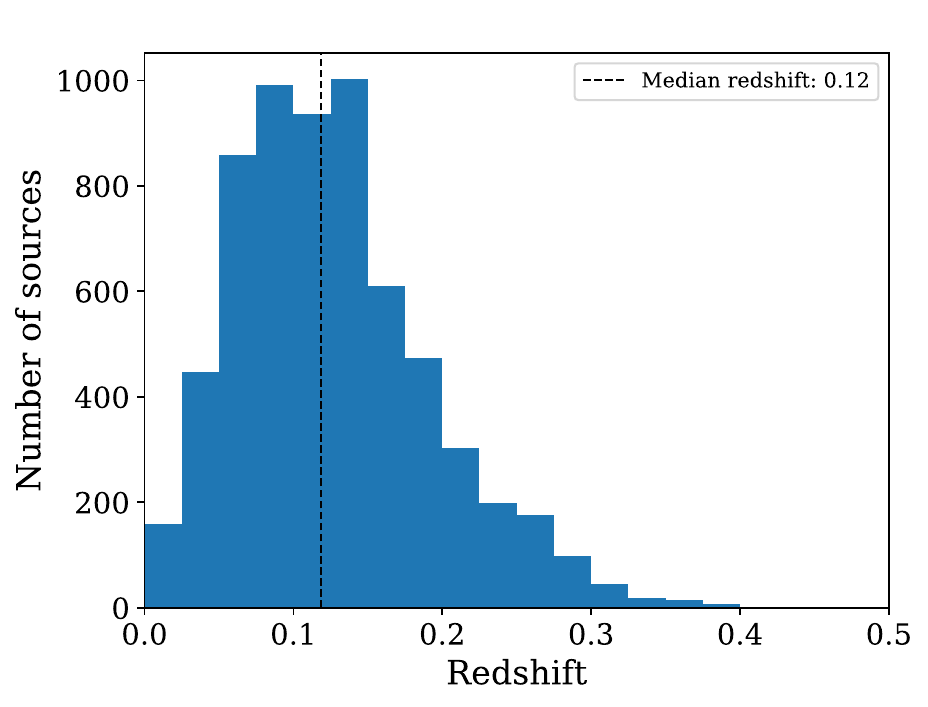}
    \caption{Left: The density of unWISE-2MASS galaxies on the sky. The gray region combines a \(\pm 10\degr\) region surrounding the galactic plane and a Planck dust map. The Large and Small Magellanic Clouds are masked as well. Right: The redshift distribution of the unWISE-2MASS catalog found by cross-matching sources with the GAMA redshift catalog. The GAMA redshifts are spectroscopically measured, so they are precise enough to neglect the statistical uncertainty. The catalog is magnitude-limited and has a tail at larger redshifts. The median redshift is 0.12, and the 10\% and 90\% percentiles are 0.05 and 0.22. \label{fig:unwise-catalog}}
\end{figure*}

The 2 Micron All Sky Survey (2MASS) is a near-infrared survey performed by a pair of ground-based near-infrared telescopes located at Mt. Hopkins, Arizona, and Cerro Tololo, Chile \citep{skrutskieTwoMicronAll2006}.
The telescopes observed the full sky in three filters: $H$, $J$, $K_s$ corresponding to central wavelengths 1.25, 1.65, and 2.16 $\mu$m.
2MASS produced an all-sky catalog containing photometry and astrometry for 471 million sources.
The Wide Field Infrared Survey (WISE) is a far-infrared space telescope which performed a survey of the entire sky in four bands, called W1, W2, W3, and W4 \citep{wrightWIDEFIELDINFRAREDSURVEY2010}.
The NEOWISE mission extends the original mission focused on near-Earth objects \citep{mainzerNEOWISEOBSERVATIONSNEAREARTH2011}.
The unWISE catalog is a reprocessing of data from the initial and extended mission, including W1 and W2 photometry from the extended mission to create the deepest WISE catalog yet \citep{schlaflyUnWISECatalogTwo2019}.
We produced galaxy overdensity maps as described in Appendix \ref{app:galaxy-map}.
We reproduce the method of \citet{kovacsStarGalaxySeparation2015} to reject sources in the \unwisetwomass catalog that are likely stellar.
We also apply magnitude cuts of $W1 \le 15.2$ and $J < 16.5$ to improve the spatial uniformity of the catalog.
The redshift distribution of this catalog is estimated by cross-matching a subset of the \unwisetwomass catalog with the Galaxy and Mass Assembly (GAMA) catalog \citep{driverGalaxyMassAssembly2022}.
Despite the filtering, there is still contamination by galactic sources near the galactic plane.
We apply a mask derived from the Planck dust map \citet{aghanimPlanck2018Results2020a} as well as excluding the sky pixels with galactic latitude (\(-10\degr<b<10\degr\)).
The Large and Small Magellanic Clouds are extended infrared sources that themselves contain stars that may be resolved by WISE and 2MASS.
Two $0.5\degr$ radius circles surrounding both the LMC and SMC are excluded.
The final galaxy overdensity map and redshift distribution are shown in Figure \ref{fig:unwise-catalog}.
The median redshift of these sources is 0.12, and the 10\% and 90\% range is 0.05 to 0.22.

\section{IceCube Data}\label{sec:IceCube}
The IceCube Neutrino Observatory is a set of 86 strings bearing 5160  digital optical modules (DOMs) \citep{hansonDesignProductionIceCube2006, aartsenIceCubeNeutrinoObservatory2017, abbasiCalibrationCharacterizationIceCube2010}.
These strings are installed 1.5 - 2.5 km below the surface of the glacial ice at the South Pole in a hexagonal array.
The DOMs carry photomultiplier tubes (PMT) and electronics to power those PMTs and communicate through the strings to the surface.
Neutrinos are not directly detected; instead, the Cherenkov radiation emitted by secondary particles is observed by the DOMs.
The timing and intensity of the light pulse in each DOM are used to reconstruct the energy and direction of the incoming neutrino.
Astrophysical muon neutrinos may produce a muon if they undergo deep inelastic scattering through the charged current channel.
This muon can traverse the detector and emit a track of Cherenkov radiation.
Muons may also be produced in the atmosphere by energetic cosmic rays.
The Earth is a natural filter for these, so a selection of Northern sky track events maximizes the neutrino sample purity.
The cosmic ray interactions in the atmosphere also produce muon neutrinos, which cannot be distinguished from astrophysical muon neutrinos on an event-by-event basis.
Based on the measured spectral shape of the diffuse muon neutrino spectrum \citep{abbasiImprovedCharacterizationAstrophysical2022a}, the astrophysical sample purity is about 1.3\%.
Although the muon energy may be estimated, the reconstructed energy is a lower bound on the neutrino energy because some energy may be lost by the muon before entering the detector.
We use a similar sample of northern-sky track-like events as in \citet{abbasiImprovedCharacterizationAstrophysical2022a}, with the restriction that we only use data from the final IceCube configuration (IC86) taken between May 13, 2011, and May 3, 2021 with three additional years of full-configuration data (IC86-2011 to IC86-2021).
This event sample contains 507017 events.
We further apply the mask described in Section \ref{sec:cat} (shown in Figure \ref{fig:unwise-catalog}) and bin the events into logarithmically space reconstructed energy bins centered at 1, 10, and 100 TeV resulting in three sets of events containing 338377, 32281, and 723 totaling 371381 events with \(2.5 > \log (E_{\rm reco}/{\rm GeV}) > 5.5\) in the unmasked sky.

The neutrino counts maps are pixelated using Healpix with \texttt{NSIDE}=128 corresponding to approximately \(0.5\degr\) pixel size in three logarithmically spaced reconstructed energy bins centered at 1, 10, and 100 TeV \citep{gorskiHEALPixFrameworkHighResolution2005}.
Each neutrino sky map is constructed similarly to the infrared galaxy map, but includes an additional weighting factor to correct the declination-dependent and energy-dependent exposure.
First, a Healpix map containing the counts, $n(\vec{x})$ with \( \vx \) being the unit vector corresponding to a point on the celestial sphere, is constructed for each energy bin, $i$.
The weight maps \( w(\vec{x}) \) are designed to make a coarse estimate of the flux incident on the detector based on the observed counts, eg., \( \phi(\vx) \approx w(\vx) n(\vx) \) where \( \phi(\vx) = \frac{dN}{dE} \) is the flux normalization assuming a power law spectrum.
A true estimate of the flux would require a more sophisticated unfolding of the detector response; however, this procedure produces an acceptance map that effectively removes the declination dependence of the effective area.
The weighted neutrino overdensity in reconstructed energy bin \(i\) is
\begin{equation}
    \label{eqn:weighted-sky-map}
    \delta_{\nu,i}(\vec{x}) = \frac{n_i(\vec{x})\,w_i(\vec{x}) - \langle n_i\, w_i\rangle}{\langle n_i\, w_i\rangle},
\end{equation}
where \(\langle n\,w\rangle\) is the weighted sky map averaged over unmasked pixels using the same mask as the \unwisetwomass overdensity.
As we only use overdensity maps, any overall normalization factor is removed in Equation \ref{eqn:weighted-sky-map}.

The weight maps are created by constructing a weighted histogram of the reconstructed sine-declination (\(\sin(\delta)\)) of simulated events.
The weights are proportional to the flux at the reconstructed energy for a power law spectrum.
We construct a Healpix map using this histogram as a lookup table for the weight at the center of the pixel.
This map is blurred with a 5\degr~Gaussian kernel to remove the discrete bands.
We do not know the spectral index of the neutrino sample a priori, so we produce weight maps for an equally spaced grid of spectral indices between \(1\le\gamma\le4\) and take the mean over the spectral index.
The weight maps are not sensitive to the choice of spectral indices and correct for the unequal sensitivity to different parts of the sky.

\section{Angular Cross Correlation} \label{sec:method}

The neutrino data are composed of three populations:
\begin{enumerate}
 \item Astrophysical neutrinos from sources correlated with the \unwisetwomass catalog. These neutrinos are produced in sources that exist in the same redshift range as the \unwisetwomass catalog.
 \item Astrophysical neutrinos uncorrelated with the galaxy catalog. These neutrinos may come from sources uncorrelated with the large-scale structure, or from outside the redshift range probed by the \unwisetwomass catalog.
 \item Atmospheric neutrinos have an anisotropic spatial distribution that is determined by the interaction of cosmic rays with the Earth's atmosphere.
\end{enumerate}
We use the parameters \(f_{\textrm{corr},i}\), \(f_{\textrm{uncorr},i}\), and \(f_{\textrm{atm},i}\) to refer to the relative contributions to the overall cross-correlation coming from each of the three groups.
If the correlated neutrinos represent a Poisson sampling of the \unwisetwomass galaxies, then the \(f\) parameters can be interpreted as the fraction of observed neutrinos belonging to each group.
The fractions add to one because they must explain all the neutrinos observed by IceCube, i.e., \(
    \label{eqn:f-sum}
    f_{\textrm{corr},i} + f_{\textrm{uncorr},i} + f_{\textrm{atm},i} = 1.
\)

\citet{fangCrosscorrelationStudyHighenergy2020} showed that the power spectrum of a multi-component population of events in reconstructed energy bin $i$ can be decomposed additively.
The  two-point angular cross-power spectrum of the two overdensity fields, as defined in Appendix~\ref{appendix:crossSpec}, can be written as
\begin{equation}
    \label{eqn:decomposition-energy-bin}
    C_{\ell\,i}^{g\nu} = f_{\textrm{corr},i}~C_{\ell \,i}^{g\nu,\,\textrm{corr}} +  f_{\textrm{atm},i}~C_{\ell\,i}^{g\nu,\,\textrm{atm}}.
\end{equation}
In this equation, we have neglected the term regarding the uncorrelated population because $\langle C_{\ell\,i}^{g\nu,\,\textrm{uncorr}}\rangle = 0$  by default. Its contribution to the covariance is minor since the covariance matrix is dominated by the atmospheric terms, as explained below. 
 
Equation \ref{eqn:decomposition-energy-bin} applies to each energy bin individually.
All the energy bins will be simultaneously used in a single likelihood to maximize the statistical significance, so we re-parameterize Equation \ref{eqn:decomposition-energy-bin}   in terms of a single correlation strength, $f_\textrm{corr}$, and spectral index, $\gamma$, assuming the astrophysical events follow a power law energy distribution:
\begin{equation}
    \label{eqn:decomposition}
    C_{\ell\,i}^{g\nu} = f_\textrm{corr}~\kappa_i(\gamma)~C_{\ell}^{gg} + f_{\textrm{atm},i}~C_{\ell\,i}^{g\nu,\,\textrm{atm}},
\end{equation}
where $\kappa(\gamma)$ is the ratio of detector acceptance in the given energy bin (\(A_i(\gamma)\)) for a given spectral index divided by the acceptance across the entire energy range (\(A(\gamma)\)) for the same spectral index, i.e., $\kappa(\gamma)={A_i(\gamma)}/{A(\gamma)}$ and 
\begin{equation}
    f_\textrm{corr} \equiv \frac{b_\nu}{b_g}\frac{n_\textrm{corr}}{n_\textrm{total}},
    \label{eqn:fcorr2ncorr}
\end{equation}
where $b_\nu$ and $b_g$ are the bias factors of the unknown neutrino source population and the galaxy sample, respectively.
Although tracers of large-scale structure can be used to infer the underlying matter density, they can be biased by the physics underlying their formation.
\citet{kaiserSpatialCorrelationsAbell1984} found that, if galaxy clusters or other large-scale structure tracers formed where the underlying density is unusually large, the correlation functions are biased by a linear factor, $b$, relative to the underlying matter correlation.
The bias can be written as
\begin{align}
    C_\ell^{gg} &= b_g^2~C_\ell^{mm},\\
    C_{\ell\,i}^{g\nu,\,\textrm{corr}} &=  b_\nu~b_g~C_\ell^{mm},\\
    C_{\ell\,i}^{\nu\nu,\,\textrm{corr}} &= b^2_\nu~C_\ell^{mm}.
\end{align}
See Appendix~\ref{appendix:crossSpec} for further discussion.
Their ratio, \(  {b_\nu}/{b_g} \), is left as a free parameter.

If the angular power spectrum is measured using the entire sky, each multipole is statistically independent, and the observed cross-power spectra follow a multivariate normal distribution. 
The use of a mask, however, causes the individual multipoles in the cross-power spectrum to be coupled. 
We may assume that the cross-power spectra follow a multivariate Gaussian distribution and use a covariance matrix to account for mode-coupling caused by the sky masking.
Specifically, we define the likelihood as
\begin{equation}
    \label{eqn:likelihood}
    \log\left(\mathcal{L}\right) = \sum_{i} -\frac{1}{2}~\vec{x}_i^T(\vec{\theta})~\Sigma_i^{-1}~\vec{x}_i(\vec{\theta}) + \textrm{constant},
\end{equation}
where  \(\vec{\theta} = (f_\textrm{corr} \gamma, f_{\textrm{atm},1}, f_{\textrm{atm},2}, f_{\textrm{atm},3})\)  are the parameters,  
\begin{equation}
    \vec{x}_i(\vec{\theta}) = C_{\ell\,i}^{g\nu} - \left(f_\textrm{corr}~\kappa_i(\gamma)~C_{\ell\,i}^{gg} + f_{\textrm{atm},i}~C_{\ell\,i}^{g\nu,\,\textrm{atm}}\right)
\end{equation}
following equation~\ref{eqn:decomposition}, and $\Sigma_i^{-1}$ is the inverse covariance matrix.

Although the atmospheric neutrino production is not actually correlated with the large-scale structure, the neutrino sample exhibits spurious correlations because we are looking at a single realization of a random process, which is itself uncorrelated with the atmospheric background on average.
At small scales, $\ell>50$, the atmospheric event distribution is smooth, so the angular correlation is zero.
We include $\ell \ge 10$.
At these large angular scales, the atmospheric events are not isotropic and will have a non-zero contribution to the cross-power spectra.
The inclusion of the $C_{\ell, i}^{g\nu, \rm atm}$ terms accounts for this anisotropic distribution of atmospheric neutrinos at low multipoles. 
The matrix multiplication in equation~\ref{eqn:likelihood} starts from a minimum multipole  \(\ell_\textrm{min}=10\). Compared to the previous work \citep{fangCrosscorrelationStudyHighenergy2020}, including information at large scales helps to optimize the sensitivity.

%



The test statistic (TS) is the log-likelihood ratio test where the null hypothesis is that \(f_\textrm{corr}=0\) and the test hypothesis is that \(f_\textrm{corr}>0\) and \(\gamma\) is a free parameter.
Details on the construction of models for \(C_{\ell\,i}^{g\nu,\,\textrm{atm}}\), the effect of the IceCube beam function, the likelihood, the test statistic, and the systematic uncertainties are described in Appendix \ref{app:statistics}.

\section{Results} \label{sec:results}

\begin{table}
    \centering
    \begin{tabular*}{\linewidth}{@{\extracolsep{\fill}} ccc }
        \hline
        & Null hypothesis & Test hypothesis \\
        \hline
        TS & - & 2.307 \\
        P-value & - & 0.23 \\
        \(\chi^2\) & 1082 & 1077 \\
        DoF & 1120 & 1122 \\
        \(n_\textrm{corr}\) & - & \(4143\) \\
        \(f_\textrm{corr}\) & - & \(0.011\) \\
        Spectral Index & - & 4.0\(^\dagger\)\\
        \(f_{\textrm{atm},1}\) & \(1.101\) & \(1.089\) \\
        \(f_{\textrm{atm},2}\) & \(1.012\) & \(1.122\) \\
        \(f_{\textrm{atm},3}\) & \(0.893\) & \(0.891\) \\
        \hline
    \end{tabular*}
    \caption{The best fit parameters found by the likelihood maximization procedure. The right column is the test hypothesis, where some neutrinos are correlated with the large-scale structure. The left column is the null hypothesis where \(f_\textrm{corr}\) is fixed to zero. The dagger (\(\dagger\)) indicates that the best fit parameter lies on the boundary of the allowed parameter space. The value for \(n_{\rm corr}\) assumes \(b_\nu=b_g\) (see Equation \ref{eqn:fcorr2ncorr}). \label{tab:best-fit-parameters}}
\end{table}

The analysis method was developed while keeping the right ascension (RA) coordinates blind, in order to avoid bias that could steer the development of the method toward a preferred outcome.
Once the method was finalized, the analysis was performed using the true RA values of the events.
The observed test statistic was \(\textrm{TS} = 2.307\).
The background test statistic distribution was numerically estimated by generating data sets containing experimental data that have had uniformly random shifts in right ascension (RA) from 0 to \(2\pi\).
The data is background-dominated, so the ``RA scrambling'' eliminates the spatial correlations while creating plausibly realistic signal-free data.
We generated 64700 RA-scrambled data sets and evaluated the test statistic for each.
The p-value (i.e., the fraction of scrambled data sets with a higher value of TS than observed for the real data) is 0.23, which is not statistically significant.
The data were well-described by the maximum likelihood test hypothesis model.
The \(\chi^2\) goodness-of-fit, defined as \(\chi^2 = \sum_{i=1}^3 \vec{x}_i^T\, \Sigma^{-1}_i\, \vec{x}_i\), was 1082 with 1120 degrees of freedom for the null hypothesis fit.
For a \(\chi^2\) distribution with \(k\) degrees of freedom, the mean is \(k\) and the standard deviation is \(\sqrt{2k}\).
The observed goodness of fit is \(-0.80\sigma\) from the mean, suggesting that the null hypothesis model describes the data well.
The fit parameters are shown in Table \ref{tab:best-fit-parameters}.

\begin{figure}
    \centering
    \plotone{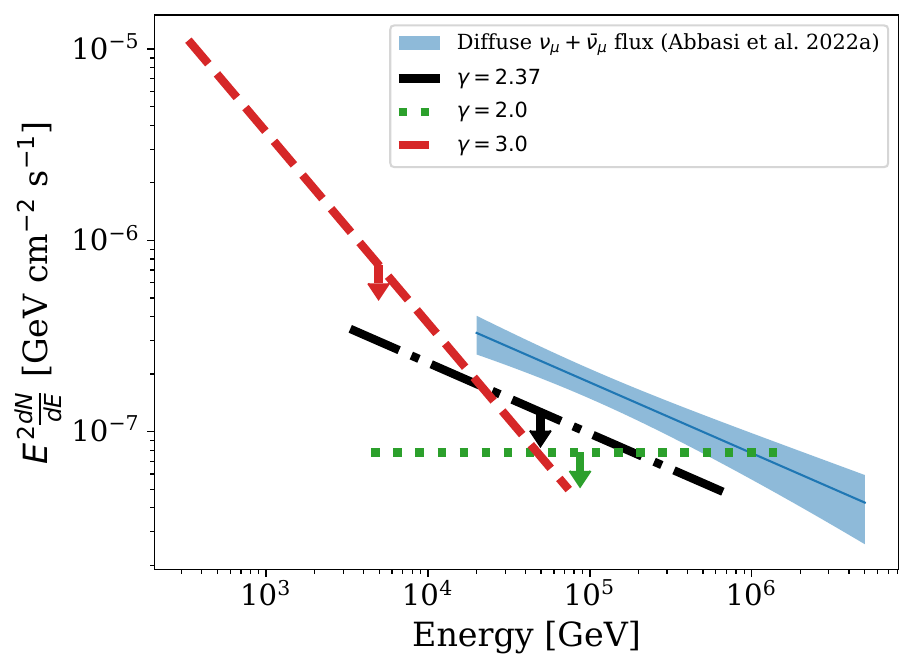}
    \caption{\label{fig:correlation-upper-limits} If the neutrinos are sampled from the unWISE-2MASS catalog, then the neutrino flux can be estimated assuming the correlation bias parameters are equal, i.e., \(b_\nu = b_g\). The derived fluxes for three different spectral indices are shown in the red dashed, black dot-dashed, and green dotted lines. The best fit diffuse flux of muon neutrinos is shown in the blue region. If the diffuse spectral index is assumed for the correlated neutrinos, the correlated flux is less than 54\% of the total diffuse flux. The cross-correlation upper limits are plotted over the 90\% sensitive energy range, which is defined as the upper and lower energy range that causes the analysis sensitivity to drop by 10\%.}
\end{figure}

The 90\% Bayesian credible upper limit for \fcorr\ is the value below which 90\% of the posterior probability lies. The posterior is proportional to the likelihood \(\mathcal{L}\) (defined through the log-likelihood in Equation \ref{eqn:likelihood}) multiplied by the prior; here we assume a flat prior on \fcorr. 
We evaluate the posterior probability distribution for \fcorr, compute its cumulative distribution function (CDF), and determine the value of \fcorr\ such that CDF(\fcorr) = 0.9. 
Practically, we calculate the CDF at evenly spaced \fcorr\ values, fit a spline to the result, and interpolate to find the \fcorr\ corresponding to 90\% enclosed probability.
When calculating an upper limit, the spectral index must be fixed.
The most natural value is the observed spectral index of diffuse astrophysical muon neutrinos: \(\gamma=2.37\) \citep{abbasiImprovedCharacterizationAstrophysical2022a}.
We also use \(\gamma=2.0\) and \(\gamma=3.0\) as generic values that span a reasonable range that is consistent with diffuse astrophysical neutrino observations.
The upper limits assume that \(b_\nu = b_g\).
The upper limits are shown in Figure \ref{fig:correlation-upper-limits}.
For \(\gamma=2.37\), less than 54\% of the astrophysical diffuse muon neutrino flux at 100 TeV as measured in \citet{abbasiImprovedCharacterizationAstrophysical2022a} can be explained by sources that are correlated with the \unwisetwomass catalog.
These results depend on the assumption that neutrinos are either correlated with the \unwisetwomass galaxies or that they are completely uncorrelated.
More sophisticated models are described in the next section.

\section{Interpretation}\label{sec:dis}
\begin{figure*}[t]
    \centering
    \plottwo{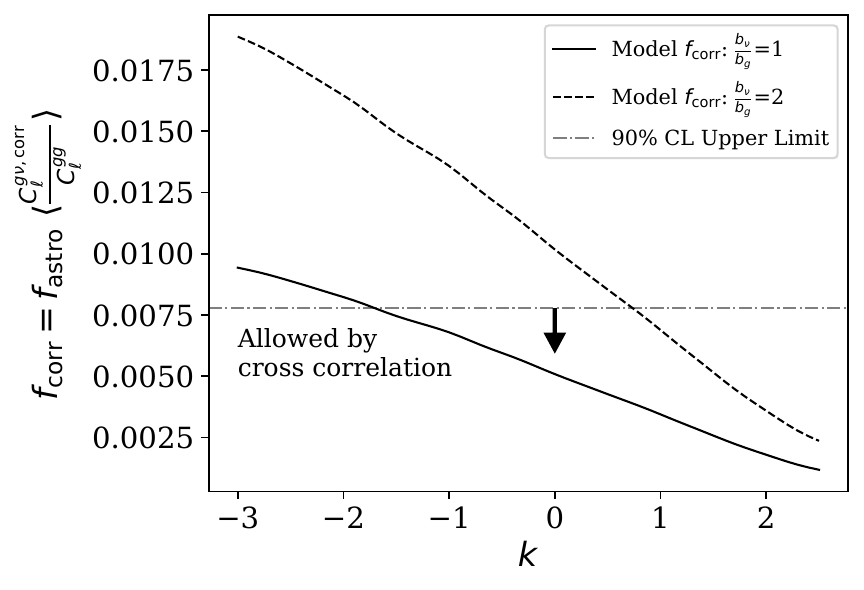}{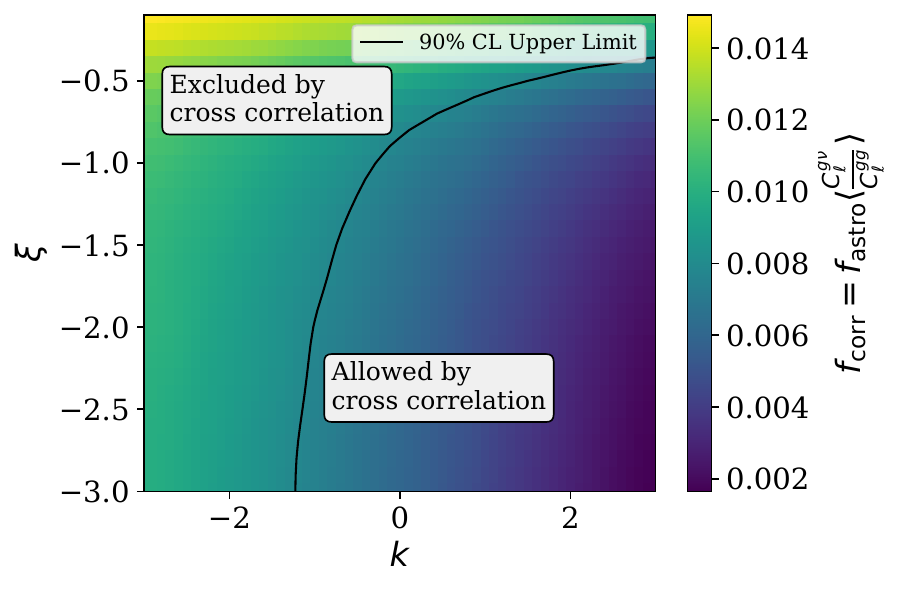}
    \caption{\label{fig:1d-source-constraints} Left: If the neutrino sources follow a distribution of the form \(\frac{dN_s}{dV}\propto (1+z)^k\), the expected correlation strength, \(f_\textrm{corr}\), can be calculated given the ratio of diffuse astrophysical neutrinos to total neutrinos, \(f_\textrm{astro}\). The expected correlation strength for equal bias parameters (solid black) and \(\frac{b_\nu}{b_g} = 2\) (dashed line) are shown. The latter case is a pessimistic model given that neutrino sources may form within the same dark matter halos as the \unwisetwomass galaxies. The actual upper limit is shown by the gray dash-dot line. The area below the line is allowed by the upper limit. Right: If the neutrino sources follow a distribution of the form  \(\frac{dN_s}{dV}=(1+z)^k~e^{z/\xi}\) and astrophysical spectral index and sample purity are equal to the best fit diffuse muon neutrino flux (\(\gamma=2.37\) and \(f_\textrm{astro}=0.013\)), the expected correlation strength can be calculated numerically using the Core Cosmology Library software. The expected correlation strength \(f_\textrm{corr}\) for equal bias parameters is shown in solid black. The region in the lower right predicts correlations that are weaker than the observed upper limit. For \(\xi \ll 1\), the limit on power law index approaches \(k\ge -1.75\) which suggests a median observed neutrino redshift \(z>0.2\).}
\end{figure*}

In the previous section, we established a limit on the flux of neutrinos correlated with the large-scale structure.
The limit on correlation strength may also be interpreted in terms of the evolution of the comoving neutrino source density with respect to redshift.
We operate under the assumption that the diffuse muon neutrino flux from \citet{abbasiImprovedCharacterizationAstrophysical2022a} is produced by sources tracing the large-scale structure over a wide redshift range, which extends well beyond what is sampled by the \unwisetwomass catalog, and that there is a single dominant neutrino source population.
The comoving density of these sources can be written as \({dN_s(z)}/{dV}\).
For a model of the comoving density, we compute the expected \(f_\textrm{corr}\) and compare this with the upper limit obtained in this work assuming the neutrino spectral energy distribution is the same as the diffuse muon neutrino result, \(\gamma=2.37\).
The calculation is done by (1) generating a Monte Carlo simulated neutrino source population, (2) calculating the neutrino flux from each source as seen at Earth, (3) constructing the distribution of neutrino flux with respect to redshift \({dN_\nu}/{dz}\), (4) calculating the expected cross-power spectrum, and (5) calculating the expected \(f_\textrm{corr}\).
Appendix \ref{app:modeling} describes the process in detail.
This approach is similar to the one in \citet{ouelletteCrosscorrelatingIceCubeNeutrinos2024b} with an additional final step that allows us to compare with our parameterization.

Several interesting neutrino source population models can be tested under this framework.
These include:
\begin{enumerate}
    \item Power law: \(\frac{dN_s}{dV}\propto (1+z)^k\)
    \item Power law with cutoff: \(\frac{dN_s}{dV}\propto (1+z)^k ~e^{-z/\xi}\)
    \item Sources tracing \unwisetwomass: \(\frac{dN_s}{dV} \propto \frac{dN_g}{dV}\)
    \item Counts tracing \unwisetwomass: \(\frac{dN_\nu}{dV} \propto \frac{dN_g}{dV}\)
    \item Sources tracing the star formation rate (SFR): \(\frac{dN_s}{dV} \propto \textrm{SFR}(z)\)
    \item Constant co-moving density: \(\frac{dN_s}{dV} \propto \textrm{constant}\)
\end{enumerate}


\textit{Power law model}: The constraints on the power law neutrino source population are shown in Figure \ref{fig:1d-source-constraints}.
We calculated the expected correlation strength for a variety of source evolution power-law indices.
Assuming equal bias parameters, we find \(k>-1.75\) at 90\% confidence.

\textit{Power law with cutoff model}: Qualitatively, this model is similar to the behavior of the star formation rate, which peaks around \(z=2\).
The jointly allowed parameter space of \(k\) and \(\xi\) is shown in Figure \ref{fig:1d-source-constraints}.
The region in the bottom right section of the figure shows the allowed values of the exponentially cutoff power-law source distribution.
Below \(\xi=-1\), the allowed values of \(k\) are not strongly affected.
Source distributions with \(\xi>-0.5\) have a higher concentration of nearby sources, so the correlation strength is inconsistent with the observed correlation upper limit.


\begin{table}
    \centering
    \begin{tabular*}{\linewidth}{@{\extracolsep{\fill}} ccc }
        \hline
        Model & \(f_\textrm{corr,model}\) & \(\frac{f_\textrm{corr,model}}{f_\textrm{corr,UL}}\)\\
        \hline
        Sources tracing \unwisetwomass & 0.016 & 2.051 \\
        Counts tracing \unwisetwomass & 0.014 & 1.846 \\
        Star formation rate & 0.002 & 0.269 \\
        Constant comoving density & 0.005 & 0.667 \\
        \hline
    \end{tabular*}
    \caption{The expected correlation strength is tabulated above for various models. We disfavor models where all the sources have the same distribution with respect to redshift as the \unwisetwomass catalog or the neutrino counts themselves follow the \unwisetwomass. We cannot rule out models where neutrino sources follow the star formation rate, or where the neutrino sources have a uniform comoving density unless the neutrino sources have a large bias parameter.\label{tab:model-correlation}}
\end{table}

\textit{Additional models}: The two models presented above contain free parameters that can be used to tune the correlation strength until it is consistent with the observed upper limit.
The following models have no free parameters other than \(b_g\) and \(b_\nu\), so we simply present the model prediction, which can be compared to the upper limit.
The model correlation strengths are summarized in Table \ref{tab:model-correlation}.
If neutrino sources trace the same redshift distribution as the unWISE-2MASS galaxies, or the neutrino sources are a subset of \unwisetwomass sources, the correlation strength is expected to be 0.016 or approximately twice the upper limit correlation.
Similarly, if the counts follow the same distribution as the \unwisetwomass sources, the correlation strength is 0.014 or 1.8 times the upper limit.
In either case, these are not consistent with the correlation non-detection.
A neutrino source population that traces the SFR produces a much weaker correlation.
The expected value is 0.002, which is approximately one-quarter of the upper limit.
Similarly, a neutrino source population that has a constant co-moving density produces a correlation of 0.005, which is two-thirds of the upper limit.

\section{Conclusions}\label{sec:conclude}

We searched for the two-point angular cross-correlation of 10 years of northern sky IceCube muon neutrinos with an infrared galaxy catalog derived from WISE and 2MASS observations.
The model for the astrophysical correlated events and atmospheric background was calculated using a forward modeling approach, which included the effect of masking and the IceCube point spread function.
The full covariance matrix was included to account for mode-coupling induced by the mask.
We maximized the likelihood of the overall correlation strength and power law spectral index across the entire energy range and the atmospheric background contribution in each reconstructed energy bin.
In the null hypothesis fit, we fit only the atmospheric background strength.
The maximum likelihood estimation yielded a test statistic of 2.307 corresponding to a significance of \(1.21\sigma\).
This is not a statistically significant result, so we placed upper limits on the correlation strength assuming several spectral indices.
If we assume that the spectral index of the diffuse muon neutrino measurement by \citet{abbasiImprovedCharacterizationAstrophysical2022a}, that the bias parameters are equal (\(b_\nu=b_g\)), and that the neutrinos are sampled from the same distribution as the unWISE-2MASS, the correlated flux is less than 54\% of the diffuse flux at 100 TeV.

We can also constrain the distribution of neutrino sources with respect to redshift by estimating the expected correlation strength under a model of neutrino source distribution with respect to redshift.
As a baseline, we select an \( (1+ z)^k\) comoving density and compute the expected correlation strength for various values of \(k\).
These model predictions are compared with the upper limit on correlation strength assuming the diffuse muon neutrino spectral index.
The upper limits disfavor models with \(k\le -1.75\).
We also tested an exponentially cutoff comoving density \( (1+ z)^k e^{z/\xi} \) and placed limits on the allowed parameter space.
Astrophysical neutrino production is often thought to trace the star formation rate.
Such a model is consistent with the upper limit on correlation strength, as are models of uniform comoving density, as long as the neutrino source bias parameters are not exceptionally large.
We disfavor models where the entire diffuse neutrino population traces the \unwisetwomass catalog as well as models where the neutrino counts follow the \unwisetwomass catalog.

Future neutrino observatories with an improved angular reconstruction and additional data will enable better measurements of anisotropies present within the diffuse neutrino flux.
Discovering the origin of the diffuse muon neutrinos may also benefit from tomographic cross-correlation studies, where the cross-correlation is performed with multiple tracers spanning a large redshift range to directly reveal the neutrino source evolution with respect to redshift.
\citet{ouelletteCrosscorrelatingIceCubeNeutrinos2024b} performed such a study; however, the use of only multipoles greater than \(\ell\ge 50\) limits the sensitivity in this work.
A tomographic cross-correlation in the style of \citet{paopiamsapConstraintsDarkMatter2024} with catalogs that include redshift information, such as the 2MASS Photometric Redshift catalog \citep{bilickiTWOMICRONALL2013} or the WISE x SuperCOSMOS Photometric redshift catalog \citep{bilickiWISESuperCOSMOSPHOTOMETRIC2016a} has the potential to reveal the redshift dependence of neutrino sources and provide insight into their identities.

\section*{Acknowledgements}
The IceCube collaboration acknowledges the significant contributions to this manuscript from David Guevel.
The authors gratefully acknowledge the support from the following agencies and institutions:
USA {\textendash} U.S. National Science Foundation-Office of Polar Programs,
U.S. National Science Foundation-Physics Division,
U.S. National Science Foundation-EPSCoR,
U.S. National Science Foundation-Office of Advanced Cyberinfrastructure,
Wisconsin Alumni Research Foundation,
Center for High Throughput Computing (CHTC) at the University of Wisconsin{\textendash}Madison,
Open Science Grid (OSG),
Partnership to Advance Throughput Computing (PATh),
Advanced Cyberinfrastructure Coordination Ecosystem: Services {\&} Support (ACCESS),
Frontera and Ranch computing project at the Texas Advanced Computing Center,
U.S. Department of Energy-National Energy Research Scientific Computing Center,
Particle astrophysics research computing center at the University of Maryland,
Institute for Cyber-Enabled Research at Michigan State University,
Astroparticle physics computational facility at Marquette University,
NVIDIA Corporation,
and Google Cloud Platform;
Belgium {\textendash} Funds for Scientific Research (FRS-FNRS and FWO),
FWO Odysseus and Big Science programmes,
and Belgian Federal Science Policy Office (Belspo);
Germany {\textendash} Bundesministerium f{\"u}r Bildung und Forschung (BMBF),
Deutsche Forschungsgemeinschaft (DFG),
Helmholtz Alliance for Astroparticle Physics (HAP),
Initiative and Networking Fund of the Helmholtz Association,
Deutsches Elektronen Synchrotron (DESY),
and High Performance Computing cluster of the RWTH Aachen;
Sweden {\textendash} Swedish Research Council,
Swedish Polar Research Secretariat,
Swedish National Infrastructure for Computing (SNIC),
and Knut and Alice Wallenberg Foundation;
European Union {\textendash} EGI Advanced Computing for research;
Australia {\textendash} Australian Research Council;
Canada {\textendash} Natural Sciences and Engineering Research Council of Canada,
Calcul Qu{\'e}bec, Compute Ontario, Canada Foundation for Innovation, WestGrid, and Digital Research Alliance of Canada;
Denmark {\textendash} Villum Fonden, Carlsberg Foundation, and European Commission;
New Zealand {\textendash} Marsden Fund;
Japan {\textendash} Japan Society for Promotion of Science (JSPS)
and Institute for Global Prominent Research (IGPR) of Chiba University;
Korea {\textendash} National Research Foundation of Korea (NRF);
Switzerland {\textendash} Swiss National Science Foundation (SNSF).

\bibliography{main}{}
\bibliographystyle{aasjournal}

\appendix
\restartappendixnumbering

\section{unWISE-2MASS Catalog Map Production} \label{app:galaxy-map}
Both the WISE and 2MASS point source catalogs contain galactic and extragalactic sources; the galactic sources must be removed to produce a pure large-scale structure tracer.
\citet{kovacsStarGalaxySeparation2015} identify a simple strategy based on color and magnitude cuts that discriminate between galaxies and stars.
The authors use the WISE source cross-matching provided with the WISE catalog to obtain J magnitudes for the WISE sources, then exclude sources with $W1 - J < -1.7$.
We also apply magnitude cuts of $W1 \le 15.2$ and $J < 16.5$ to improve the spatial uniformity of the catalog.
Even with these cuts, the stellar contamination near the galactic plane remains high, so the authors apply a galactic plane mask derived from dust emission from \citet{schlegelMapsDustInfrared1998}.
The authors estimate the completeness of the catalog by identifying a subset of sources with spectroscopically confirmed classifications as galaxies or stars.

The authors find that the catalog contains 70\% of the spectroscopically confirmed galaxies with only 1.2\% contamination by stars.
We reproduce these steps to produce an updated catalog using the additional WISE data in the unWISE catalog.
We cross-match the unWISE sources with the 2MASS Point Source Catalog with a maximum tolerance of 3\arcsec.
The resulting galaxy catalog contains 2.4 million sources distributed across 21,200 square degrees or 51\% of the sky.
The distribution of unWISE-2MASS galaxies is shown in Figure \ref{fig:unwise-catalog}.
Despite the filtering, there is still contamination by galactic sources near the galactic plane.
The galactic plane is masked using a mask derived from the Planck dust map \citep{aghanimPlanck2018Results2020a} as well as excluding the sky pixels with galactic latitude (\(-10\degr<b<10\degr\)).
The Large and Small Magellanic Clouds are extended infrared sources that themselves contain stars that may be resolved by WISE and 2MASS.
Two $0.5\degr$ radius circles surrounding both the LMC and SMC are excluded.
The \unwisetwomass catalog is pixelated to a Healpix map in equatorial coordinates with NSIDE equal to 128.
This degree of pixelation corresponds to approximately a \(0.5\degr\) pixel size.
The IceCube point spread function has a full-width half-max greater than this, so there is no benefit to finer pixelation.
We construct a galaxy overdensity map, defined as 
\begin{equation}
    \delta_g(\vec{x}) = \frac{n(\vec{x}) - \langle n \rangle}{\langle n \rangle},
\end{equation}
where \(\langle n \rangle\) is the average over the unmasked pixels and \(\vec{x}\) is a unit vector.

We estimate the redshift distribution of the galaxies by cross-matching the unWISE-2MASS catalog with another galaxy catalog featuring spectroscopically measured redshifts.
The Galaxy and MASS Assembly (GAMA) survey \citep{driverGalaxyMassAssembly2022} contains spectroscopic redshifts for galaxies spanning approximately 250 square degrees.
For each source in the unWISE-2MASS catalog, we search for a GAMA source within 3\arcsec. 
If there is one, we assign the GAMA source's redshift to the unWISE-2MASS source.
We find 6349 likely matches.
The distribution of these redshifts is shown in Figure \ref{fig:unwise-catalog}.

\section{Cross power spectrum and bias factors}\label{appendix:crossSpec}

The two-point angular cross-power spectrum of two overdensity fields is defined by
\begin{equation}
    C_{\ell\,i}^{g\nu} = \frac{4\pi}{2\ell + 1}\int d\cos(\theta) ~\langle \delta_g(\vec{x})~\delta_{\nu,i}(\vec{x}')\rangle ~P_\ell^*(\cos(\theta)),
\end{equation}
where \(\cos(\theta)=\vec{x}\cdot\vec{x}'\) for unit vectors \(\vec{x}\) and \(\vec{x}'\), \(P_\ell^*(\cos(\theta))\) are Legendre polynomials, and the angular brackets indicate the average over field configurations.
If \(C_{\ell\,i}^{g\nu}\) is greater than zero, there is some common anisotropy shared by the overdensity fields at an angular scale \(\ell\).
The overdensity maps for galaxies or neutrinos can be represented by a decomposition into multipoles
\begin{equation}
    \delta(\vec{x}) = \sum_{\ell=0}^\infty \sum_{m=-\ell}^\ell a_{\ell m} Y_{\ell m}(\vec{x}).
\end{equation}
The estimator for the cross-power spectrum over a partially masked sky given the spherical harmonic decomposition is
\begin{equation}
    C_{\ell\,i}^{g\nu} = \frac{1}{(2\ell + 1)f_\textrm{sky}}\sum_{\ell=-m}^m a_{\ell m}^{g^*} {a_{\ell m,i}^{\nu}}.
\end{equation}

The bias factors \(b_\nu\) and \(b_g\) describe how strongly clustered the large-scale structure tracers are relative to the underlying matter density.
The clustering is influenced by the particular physics of the formation of the tracer.
At large scales, the bias parameters are linearly proportional to the angular power spectrum of the matter density.
These factors can be written as
\begin{align}
    \label{eqn:bias-factors}
    C_\ell^{gg} &= b_g^2~C_\ell^{mm},\\
    \label{eqn:bias-factors2}
    C_{\ell\,i}^{g\nu,\,\textrm{corr}} &=  b_\nu~b_g~C_\ell^{mm},\\
        \label{eqn:bias-factors3}
    C_{\ell\,i}^{\nu\nu,\,\textrm{corr}} &= b^2_\nu~C_\ell^{mm}.
\end{align}
Substituting Equation \ref{eqn:bias-factors} into Equation \ref{eqn:bias-factors2} finds that 
\begin{equation}
    \label{eqn:gn-to-gg}
    C_{\ell\,i}^{g\nu,\,\textrm{corr}} =  \frac{b_\nu}{b_g}C_\ell^{gg}.
\end{equation}

We assume that the bias factors do not depend on neutrino energy, so that
Equations \ref{eqn:bias-factors} and \ref{eqn:bias-factors2} remain true for any energy binning scheme.
The values of the bias parameters depend on the physics of the tracer formation, but their values are typically $\mathcal{O}(1)$.
The bias factors can usually be estimated by comparing the autocorrelation with simulations of the underlying matter density; however, the neutrino autocorrelation is too noisy to do this due to the low astrophysical sample purity.

\section{Details of Statistical Analysis} \label{app:statistics}

Equation \ref{eqn:decomposition} describes the true cross-power spectrum; however, the finite beam size and partial sky visibility must be accounted for.
We adopt a forward-modeling approach in the statistical evaluation of the cross-power spectrum; thus, all the cross-power spectra referred to below are to the \textit{observed} quantities, which are modulated by the point spread function and mask rather than the \textit{true} quantities.

The mapping between the true cross-power spectrum and the observed cross-power spectrum would typically be described by    $C_\ell^{g\nu, \textrm{corr}}=B^g_\ell\,B^\nu_\ell\,C_\ell^{g\nu, \textrm{true}}$
where \(B^g_\ell \) and \(B^\nu_\ell \) are the beam functions of the galaxy and neutrino maps.
This is only true if the instrument PSF is radially symmetric about the source and the same in every direction; neither of these conditions are met by IceCube.
We directly estimate the observed \(C_\ell^{g\nu,\, \textrm{corr}} \) in each energy bin through Monte Carlo simulation.
We use detailed simulations of the detector response to an impulse flux, which can be reweighted to match whatever spatial distribution and energy spectrum is desired.
We use these simulations to construct 1000 purely astrophysical neutrino data sets, which are Poisson sampled from the unWISE-2MASS template, including the effects of the point spread function and sky pixelation.
In these simulations, we use a neutrino energy spectrum following a power law with an index equal to 2.5.
The choice of energy spectral index in the model construction has a negligible effect.

A model for the atmospheric contribution to the cross-correlation, \(C_\ell^{g\nu,\, \text{atm}} \) is also necessary.
The atmospheric model in each bin is generated similarly to the signal model; however, it is generated in two ways to serve as a cross-check on the consistency of the model.
The first exploits the fact that the IceCube data is background-dominated.
The data are binned in equal-sized bins of sine declination, and a Gaussian kernel density estimate is constructed to form a continuous function as a function of declination.
The value of this function of declination is then applied to every pixel in a Healpix map, and that map is cross-correlated with the \unwisetwomass map.
The resulting cross-spectrum is used as a template for the atmospheric neutrino contribution to the total power spectrum.
The aggregation of data over sine declination almost completely destroys the spatial correlation in the data while reducing exposure to detector systematic uncertainty.
In our second approach, the model was estimated using Monte Carlo data.
Simulated atmospheric background events were generated and processed following the same kernel density estimation procedure.
The differences between the models were negligible, suggesting that the systematic difference between these models is small.

\begin{figure}
    \centering
    \plotone{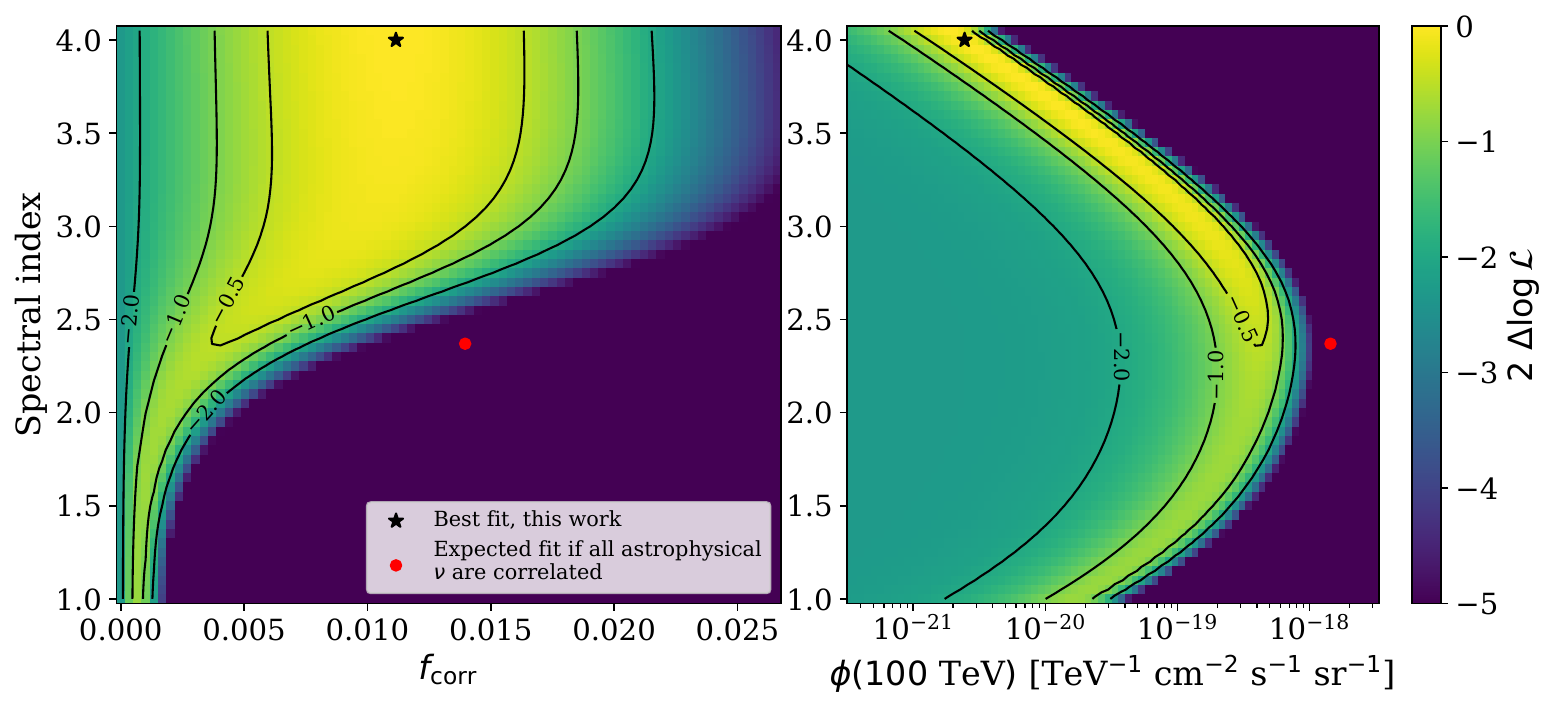}
    \caption{\label{fig:2d-likelihood-scan} Left: The change in likelihood relative to the maximum likelihood value for the two astrophysical parameters is shown in color and with contours. The maximum likelihood value is on the boundary with \(\gamma=4.0\). The test statistic is not statistically significant for the maximum likelihood parameter values. The expected value of \(f_\textrm{corr}\) is given if all of the astrophysical neutrinos were correlated with the \unwisetwomass catalog assuming the spectral index is the same as the diffuse measurement \citep{abbasiImprovedCharacterizationAstrophysical2022a}. Right: The same likelihood is shown in terms of the differential flux at 100 TeV.}
\end{figure}

The cross-power spectra are normally distributed as written in Equation \ref{eqn:likelihood}.
Ideally, each multipole in the cross-power spectrum is distributed independently; however, using a mask creates correlations between the multipoles.
Because the data is dominated by atmospheric events, the covariance matrix is estimated by creating 64700 sets of atmospheric-only data.
The cross power spectra are computed for each energy bin in each trial, and a covariance matrix, \(\Sigma_{i,{\ell_1} \ell_2}=\textrm{Cov}[C^{g\nu}_{{\ell_1}\,i}, C^{g\nu}_{{\ell_2}\,i}] \) is constructed for each energy bin from these pseudo-trial power spectra.
Energy bins are assumed to be independent: \(\textrm{Cov}[C^{g\nu}_{{\ell_1}\,i}, C^{g\nu}_{{\ell_2}\,j}] = 0 \) for \(i\ne j\).
The change in likelihood for different parameter values is shown in Figure \ref{fig:2d-likelihood-scan}.

\begin{figure}
    \centering
    \plotone{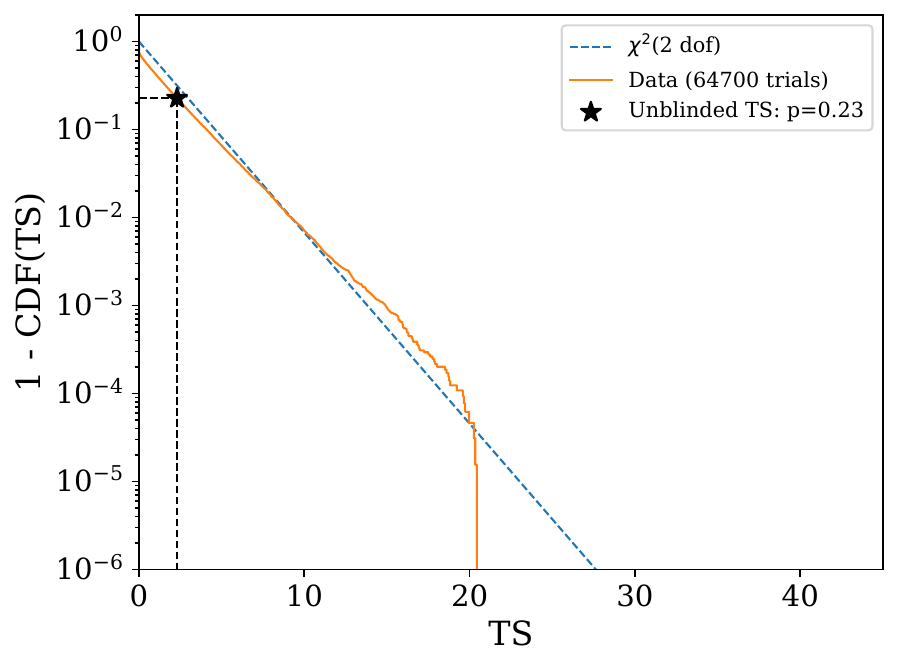}
    \caption{\label{fig:ts-distribution} The TS distribution for RA-scrambled synthetic data sets is shown above. The blue dotted line shows a \(\chi^2\) distribution with two degrees of freedom. The unblinded TS is shown with a black star. The unblinded TS is 2.307, which is not statistically significant.}
\end{figure}

We evaluate the statistical significance using a log-likelihood ratio test.
The null hypothesis is that the data sample does not contain a population of neutrinos correlated with the \unwisetwomass galaxies, \(f_\textrm{corr}=0\), and the test hypothesis is that \(f_\textrm{corr}>0\).
The test statistic can be written
\begin{equation}
    \begin{split}
    \textrm{TS} =& ~2~(\log(\mathcal{L}(\hat{f}_\textrm{corr}, \hat{\gamma}, \hat{f}_{\textrm{atm},1}, \hat{f}_{\textrm{atm},2}, \hat{f}_{\textrm{atm},3}) \\
    & - \log(\mathcal{L}(0, \hat{\gamma}, \hat{f}_{\textrm{atm},1}, \hat{f}_{\textrm{atm},2}, \hat{f}_{\textrm{atm},3})),
    \end{split}
\end{equation}
where the likelihoods in the numerator and denominator are independently maximized with respect to the parameters.
Although these are nested hypotheses, we constrain the \(f_\textrm{corr}\) to be greater than or equal to zero.
The null hypothesis lies on the boundary of the test hypothesis, so Wilk's theorem cannot be used to estimate the null hypothesis test statistic distribution.
We numerically estimate the null hypothesis test statistic distribution by simulating 64700 pseudo-trials with atmospheric neutrinos and computing the test statistic for each pseudo-trial.
The resulting test statistic distribution resembles a \(\chi^2\) distribution with two degrees of freedom; however, half of the trials have \(\textrm{TS}=0\).
This numerical test statistic distribution is used to evaluate the statistical significance of the likelihood ratio.
The test statistic distribution is shown in Figure \ref{fig:ts-distribution}.

Systematic uncertainty arises from various unmodeled or mismodeled effects that can affect the inference process.
Often, systematic uncertainty can be mitigated using data-driven methods.
For example, the atmospheric model used in the cross-correlation has been compared with an atmospheric model generated using simulated events.
The resulting models are consistent with the expected statistical uncertainty of the cross-correlation.
The systematic effect of the detector calibration on the entire analysis pipeline was evaluated using simulated data sets.
The simulated data sets were created with the nominal simulated data set, and the fit was performed with models generated with simulated data sets with detector configuration and ice properties varied.
We varied the DOM efficiency (\(\pm10\%\))), hole ice scattering, ice scattering (\(\pm 5\%\))), and ice absorption (\(\pm 5\%\))).
The systematic parameters that are varied are summarized in Table \ref{tab:systematics-mc-sets}.
Two tests were run.
The first used simulated data sets with only atmospheric neutrinos.
The second used a realistic level of signal injection, about 2500 correlated neutrinos.
The systematic uncertainty is defined as \(\sigma_{f,\textrm{sys}} = \max_i(|f_i - f_\textrm{nominal}|)\)
and \(\sigma_{\gamma,\textrm{sys}} = \max(|\gamma_i - \gamma_\textrm{nominal}|)\)
where \(i\) indexes the set of MC systematic data sets.
The average fractional systematic uncertainty on the correlation strength is approximately 20\%, and the average fractional systematic uncertainty on the spectral index is approximately 5\%.

\begin{table}
    \label{tab:systematics-mc-sets}
    \centering
    \begin{tabular*}{\linewidth}{@{\extracolsep{\fill}} ccc }
        \hline
        Systematic varied & Value & Description \\
        \hline\rule{0pt}{2.5ex}
        DOM Efficiency & \(\pm 10\%\) & Ratio observed and simulated DOM photoelectrons \\
        Hole Ice P0 & \(\pm 1\) & Hole ice scattering parameter \\
        Ice scattering & \(\pm 5\%\) & Scattering of photons in ice \\
        Ice absorption & \(\pm 5\%\) & Scattering of photons in ice \\
        \hline
    \end{tabular*}
    \caption{Systematic variations in these parameters were considered while estimating systematic uncertainty in the fit results.}
\end{table}

\section{Cross-Correlation Modeling} \label{app:modeling}
For each of the models listed in Section \ref{sec:dis}, we generate ensembles of neutrino sources using the FIRst Extragalactic Simulation Of Neutrinos and Gamma-rays (FIRESONG) \citep{tungFIRESONGPythonPackage2021}.
FIRESONG is a Python code designed to simulate realistic neutrino source distributions and calculate the fluxes of those individual neutrino sources as seen on Earth.
FIRESONG constructs a probability density with respect to redshift, taking into account cosmological evolution and the evolution of the source population with respect to redshift.
This probability density is sampled, and the number of samples is fixed such that the expected flux from all the sources is equal to the observed diffuse muon neutrino flux.
The flux of individual sources includes the effects of intrinsic luminosity functions, power-law spectra, the cosmological dimming that causes high-energy events to redshift out of the IceCube sensitive energy range, and the luminosity distance.
The flux from a neutrino source with luminosity \(L\) and redshift \(z\) is \(F = \frac{L}{D_L^2 (1+z)^\gamma}\) where \(D_L\) is the luminosity distance and the additional factor accounts for the cosmological redshift of a power law spectrum.
Although FIRESONG produces fluxes, the cross-correlation is calculated using neutrino counts.
The count rate observed from an individual source is linearly proportional to the flux observed from that source, so the distribution of flux with respect to redshift is the same as the distribution of the observed counts with respect to redshift after normalizing the distributions.
The weighting applied to the neutrino overdensity maps ensures that the spatially dependent effective area has a minimal effect.

After constructing a model for the distribution of neutrino flux \(\frac{d\phi}{dz}\) (or equivalently counts \(\frac{dN_\nu}{dz}\)) as a function of redshift, we use the Core Cosmology Library (CCL) to model the cross-power spectrum \citep{chisariCoreCosmologyLibrary2019}.
The cross-correlation strength due to astrophysical neutrinos is
\begin{equation}
    \label{eqn:cross-correlation-model}
    C_\ell^{g\nu,\,\textrm{corr}} = b_g ~b_\nu \int d\chi ~\chi^{-2}~ \lambda_\nu(\chi)~\lambda_g(\chi)~P_k\left(\frac{\ell+\frac{1}{2}}{\chi}\right),
\end{equation}
where \( \lambda_g = \frac{1}{N_g}\frac{dN_g}{dz(\chi)}\), \(N_g\) is the number of sources in the galaxy catalog, \(\lambda_\nu = \frac{1}{N_\nu}\frac{dN_\nu}{dz(\chi)}\) is the distribution of neutrino counts, \(\chi\) is the co-moving distance, and \(P_k\) is the matter density power spectrum.
Similarly, the galaxy autocorrelation can be written as
\begin{equation}
    \label{eqn:auto-correlation-model}
    C_\ell^{gg} = b_g^2 \int d\chi ~\chi^{-2}~ \lambda_g(\chi)^2 ~P_k\left(\frac{\ell+\frac{1}{2}}{\chi}\right).
\end{equation}
The galaxy distribution \(\lambda_g\) has been obtained by cross-matching a subset of the \unwisetwomass catalog with the GAMA spectroscopic survey.
This distribution is shown in Figure \ref{fig:unwise-catalog}.
CCL is used to evaluate the integrals in Equations \ref{eqn:cross-correlation-model} and \ref{eqn:auto-correlation-model}.

As we have defined the likelihood and Equation \ref{eqn:gn-to-gg}, the correlation strength \(f_\textrm{corr}\) is equivalent to the ratio of the cross-power spectra and galaxy autocorrelation averaged over multipoles, 
\begin{equation}
    \label{eqn:correlation-ratio}
    f_\textrm{corr} = \frac{n_{\rm corr}}{n_{\rm total}}\frac{b_\nu}{b_g} \approx f_{\rm astro}\biggr\langle\frac{C_\ell^{g\nu,\,\textrm{corr}}}{C_\ell^{gg}}\biggr\rangle.
\end{equation}
Inspection of equations \ref{eqn:cross-correlation-model} and \ref{eqn:auto-correlation-model} shows the equality on the left is exactly when \(\lambda_g = \lambda_\nu\).
This remains approximately true as long as the redshift range of the galaxy catalog is not too large.
We have verified numerically that the ratio of the cross-power spectrum and galaxy autocorrelation vary only by a few percent, dependent on multipole \(\ell\).
The factor \(f_\textrm{astro}\) here represents the astrophysical sample purity for all neutrinos regardless of energy.
This factor can be estimated using the best-fit power law diffuse muon neutrino single power law parameters using only the events with reconstructed energy in the range used in this analysis.
In this case, \(f_\textrm{astro} =\frac{n_\textrm{astro}}{n_\textrm{total}} = 0.013\).

\end{document}